\documentclass[aps,prb,floatfix,twocolumn,showpacs]{revtex4}%
\usepackage{amsmath}
\usepackage{graphicx}
\usepackage{amsfonts}
\usepackage{amssymb}%
\begin{document}
\title{Magnetoresistance of a quantum dot with spin-active interfaces}
\author{Audrey Cottet}
\email{cottet@lps.u-psud.fr}
\affiliation{Laboratoire de Physique des Solides, Universit\'{e} Paris-Sud, 91405 Orsay, France}
\author{Mahn-Soo Choi}
\email{choims@korea.ac.kr}
\affiliation{Department of Physics and Astronomy, University of Basel, Klingelbergstrasse
82, 4056 Basel, Switzerland}

\begin{abstract}
We study the zero-bias magnetoresistance ($\mathrm{MR}$) of an interacting
quantum dot connected to two ferromagnetic leads and capacitively coupled to a
gate voltage source $V_{g}$. We investigate the effects of the spin-activity
of the contacts between the dot and the leads by introducing an effective
exchange field in an Anderson model. This spin-activity makes easier negative
$\mathrm{MR}$ effects, and can even lead to a giant $\mathrm{MR}$\ effect with
a sign tunable with $V_{g}$. Assuming a twofold orbital degeneracy, our
approach allows to interpret in an interacting picture the $\mathrm{MR}%
(V_{g})$ measured by S. Sahoo\textit{\ et al. }[Nature Phys. \textbf{2}, 99
(2005)] in single wall carbon nanotubes with ferromagnetic contacts. If this
experiment is repeated on a larger $V_{g}-$range, we expect that the
$\mathrm{MR}(V_{g})$ oscillations are not regular like in the presently
available data, due to Coulomb interactions.

\end{abstract}
\pacs{73.23.-b, 75.75.+a, 85.75.-d}
\date{\today}
\maketitle

\section{Introduction}

\label{sec:introduction}

The quantum mechanical spin degree of freedom is now widely exploited to
control current transport in electronic devices. For instance, the readout of
magnetic hard disks is based on the spin-valve effect, i.e. the tunability of
a conductance through the relative orientation of some ferromagnetic
polarizations \cite{Prinz}. However, realizing spin injection in mesoscopic
conductors would allow to implement further functionalities, like e.g. a gate
control of the spin valve effect \cite{Datta,Schapers}. Importantly,
electronic interaction effects can occur in mesoscopic structures, due to the
electronic confinement. This raises the fundamental question of the interplay
between spin-dependent transport and electronic interactions.

Upon scattering on the interface between a ferromagnet (F) and a non-magnetic
material, electrons with spin parallel or antiparallel to the magnetization of
F can pick up different phase shifts, because they are affected by different
scattering potentials. This \textit{Spin-Dependence} of Interfacial Phase
Shifts (SDIPS) can modify significantly the behavior of mesoscopic circuits.
First, when a mesoscopic conductor is connected to several F leads with non
collinear polarizations, the SDIPS produces an interfacial precession of spins
which can modify current transport in the device
\cite{FNF,Luttinger,Wetzels,Ciuti,ReviewBraatas}. Secondly, in collinear
configurations, precession effects are not relevant, but the SDIPS can modify
mesoscopic coherence effects. For instance, in superconducting/ferromagnetic
hybrid circuits, the SDIPS introduces a phase shift between electron and holes
correlated by Andreev reflection \cite{SF}. References \onlinecite{Tokuyasu}
and \onlinecite{SF2005} have identified signatures of this effect in the
experiments of Refs. \onlinecite{Tedrow} and \onlinecite{Takis}, respectively.
In principle, normal systems in collinear configurations can also be affected
by the SDIPS. Indeed, from Ref. \onlinecite{wire2005}, the SDIPS should
produce a spin-splitting of the resonant states in a ballistic interactionless
wire contacted with collinearly polarized ferromagnetic leads. However, this
has not been confirmed experimentally yet \cite{comment}. \begin{figure}[ptb]
\centering\includegraphics[width=0.55\linewidth]{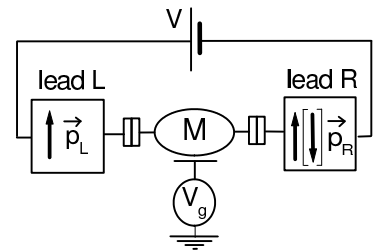}\caption{Mesoscopic
element M connected to ferromagnetic leads $L$ and $R$. The magnetic
polarizations $\vec{p}_{L}$ and $\vec{p}_{R}$ of leads $L$ and $R$ can be
parallel (configuration P) or antiparallel (configuration AP). The element M
is capacitively coupled to a gate voltage source $V_{g}$.}%
\label{Figure1}%
\end{figure}

Recently, Ref. \onlinecite{Sahoo} has reported current measurements in a
single wall carbon nanotube (SWNT) connected to two ferromagnetic leads with
collinear polarizations. The asymmetries observed in the magnetoresistance
($\mathrm{MR}$) of the SWNT versus gate voltage are strikingly similar to
those predicted by Ref. \onlinecite{wire2005} for an interactionless wire
subject to the SDIPS\cite{question}. However, the SWNT of Ref. \onlinecite
{Sahoo} showed a quantum dot behavior with strong Coulomb Blockade effects, as
demonstrated in a great number of experiments with non-magnetic leads (see
e.g. \cite{expCB,Sapmaz}).\ Therefore, one important question is how
interaction effects modify the scheme proposed by Ref. \onlinecite
{wire2005}.\ The problem of the effects of interactions on the transport
properties of a central region connected to ferromagnetic contacts has already
been considered in various regimes, like e.g. the Coulomb blockade
regime\cite{Wetzels, CB,Braun,Weymann}, the Kondo regime\cite{Kondo,Martinek},
the Luttinger liquid regime\cite{Luttinger,Luttinger2} and the marginal Fermi
liquid regime\cite{MFL}. This article develops an approach suitable for the
limit of Ref. \onlinecite{Sahoo} and studies, for the first time, the effect
of the SDIPS on a quantum dot. We consider a quantum dot coupled to metallic
leads through spin-active interfaces.\textbf{\ }We use an Anderson model to
study the $\mathrm{MR}$ of the circuit above the Kondo temperature, but beyond
the sequential tunneling limit. The SDIPS is taken into account through an
effective spin-splitting of the dot energy levels. This splitting makes easier
negative $\mathrm{MR}$ effects. When it is strong enough, it can even lead to
a giant $\mathrm{MR}$ with a sign oscillating with the dot gate voltage
$V_{g}$, similarly to what has been found in the non-interacting case. In the
non-interacting case, assuming that the properties of the contact are constant
with energy and that the SDIPS is too weak to split the conductance peaks, one
finds that the $\mathrm{MR}(V_{g})$ pattern is similar for all conductance
peaks. In contrast, the effect of the SDIPS depends on the occupation of the
dot in the interacting case. This is in apparent contradiction with the data
of Ref. \onlinecite{Sahoo} because, in the $V_{g}-$range presented in this
Ref., the $\mathrm{MR}(V_{g})$ oscillations are regular. Using a two-orbitals
model, which takes into account the $K-K^{\prime}$ orbital degeneracy commonly
observed for SWNTs (see e.g. Refs. \onlinecite
{Liang,Jarillo, Sapmaz, Moriyama, Babic, Babic2}), one can solve this
discrepancy. In this framework, we expect non-regular $\mathrm{MR}(V_{g})$
oscillations if the experiment is repeated on a larger $V_{g}-$range.

This article is organized as follows: we start with summarizing the results
found for the non-interacting case in section~\ref{sec:non-interacting}. Then
we introduce a model for the interacting case in section~\ref{sec:interacting}%
. Section~\ref{sec:single-level} addresses the case of a one-orbital quantum
dot circuit, and section~\ref{sec:two-level} the case of a
two-degenerate-orbitals quantum dot circuit. Finally
section~\ref{sec:conclusion} concludes.

\section{Model}

\label{sec:model}

We consider a mesoscopic element M connected to ferromagnetic leads $L$ and
$R$ (Fig. \ref{Figure1}). The chemical potential of M can be shifted by
$e\alpha V_{g}$ using the gate voltage $V_{g}$, with $\alpha$ the ratio
between the gate capacitance and the total capacitance of $M$. The magnetic
polarizations $\vec{p}_{L}$ and $\vec{p}_{R}$ of leads $L$ and $R$ can be
parallel (configuration $c=P$) or antiparallel (configuration $c=$ $AP$).

\subsection{Non-interacting case}

\label{sec:non-interacting}

Before introducing the interacting model investigated in this article, it is
useful to reconsider the results obtained by Ref. \onlinecite{wire2005} for
the case in which M is a non-interacting single-channel ballistic wire of
length $L$. In a scattering approach\cite{Blanter}, the conductance of the
circuit depends on the transmission probability $T_{l}^{c,\sigma}$ for
electrons with spin $\sigma\in\{\uparrow,\downarrow\}$ through contact
$l\in\{L,R\}$, and on the reflection phase $\varphi_{l}^{c,\sigma}$ for
electrons with spin $\sigma$ coming from the wire towards contact $l$. The
index $c=P[AP]$ denotes the parallel [antiparallel] leads configuration. A
spin dependence of $\varphi_{l}^{c,\sigma}$ can occur due to the magnetic
properties of the contact materials used to engineer lead $l$. Due to size
quantization, the conductance $G^{c}(V_{g})$ of the circuit in configuration
$c$ presents Fabry-Perot like resonances for $E_{d,\sigma}^{c}\sim0$, with
$E_{d,\sigma}^{c}=(2\pi d-\varphi_{L}^{c,\sigma}-\varphi_{R}^{c,\sigma})(2\pi
N_{F}^{M})^{-1}-e\alpha V_{g}-E_{F}^{M}$ a resonant energy,\ $d\in\mathbb{Z}$,
$E_{F}^{M}$ the wire Fermi energy, $N_{F}^{M}$ the density of orbitals states
at the Fermi level in the wire and $\overline{\sigma}$ the spin direction
opposite to $\sigma$ (we have used $e\alpha V_{g}\ll E_{F}^{M}$). From this
Eq., in configuration $c$, the SDIPS produces a spin-splitting
\begin{equation}
g\mu_{B}h_{SDIPS}^{c}=E_{d,\downarrow}^{c}-E_{d,\uparrow}^{c}=\sum
_{l\in\{L,R\}}\frac{\varphi_{l}^{c,\uparrow}-\varphi_{l}^{c,\downarrow}}{\pi
N_{F}^{M}} \label{split}%
\end{equation}
of the resonant energies. When the effective field $h_{SDIPS}^{c}$ is strong
enough to produce a spin-splitting of the conductance peaks, the circuit can
display a giant $\mathrm{MR}$ effect with a sign oscillating with $V_{g}$, due
to the strong shift of the conductance peaks from the $P$ to the $AP$
configurations. In the opposite case, $\mathrm{MR}$ remains smaller, but the
SDIPS can still be detected through characteristic asymmetries in the
oscillations of $\mathrm{MR}$ versus $V_{g}$ (see Fig. 2-right of Ref.
\onlinecite{wire2005}). Importantly, assuming that $T_{l}^{c,\sigma}$ and
$\varphi_{l}^{c,\sigma}$ are constant with $V_{g}$, one has $G^{c}%
(V_{g})=G^{c}(V_{g}+[2/e\alpha N_{F}^{M}])$ . This implies that when
$h_{SDIPS}^{c}$ is not strong enough to produce a spin-splitting of the
conductance peaks, the $\mathrm{MR}(V_{g})$ pattern is similar for all the
peaks displayed by $G^{P}(V_{g})$.

\subsection{Interacting case}

\label{sec:interacting}

We now assume the presence of strong Coulomb interactions inside M, such that
we have a quantum dot connected to ferromagnetic leads. Such a system can be
realized for instance by using granular films\cite{granular},
nanoparticles\cite{nanoparticles}, carbon nanotubes\cite{Sahoo,tubes}, or
$C_{60}$ molecules\cite{molecules}. In the non-interacting case of
section~\ref{sec:non-interacting}, we have considered that the spin-dependent
confinement potential felt by electrons causes the SDIPS, which leads to the
spin-splitting of the resonant states. In the interacting case, the scattering
approach is not suitable anymore. However, the energy of the quasi-bound
single particle states in quantum dot M can depend on spin due to the
spin-dependent confinement potential. On this ground, we adopt the effective
Anderson hamiltonian
\begin{equation}
H=H_{dot}+H_{leads}+H_{c} \label{H}%
\end{equation}
with
\begin{align*}
H_{dot}  &  =\sum\limits_{d,\sigma}\xi_{d\sigma}c_{d\sigma}^{\dagger
}c_{d\sigma}+\sum\limits_{\substack{d,d^{\prime},\sigma,\sigma^{\prime}
\\(d,\sigma)\neq(d^{\prime},\sigma^{\prime})}}\frac{U}{2}n_{d\sigma
}n_{d^{\prime}\sigma^{\prime}}\\
H_{leads}  &  =\sum\limits_{k,\sigma}\xi_{k\sigma}c_{k\sigma}^{\dagger
}c_{k\sigma}\\
H_{c}  &  =\sum\limits_{d,k,\sigma}\left(  t_{d\sigma}^{k}c_{d\sigma}%
^{\dagger}c_{k\sigma}+(t_{d\sigma}^{k})^{\ast}c_{k\sigma}^{\dagger}c_{d\sigma
}\right)
\end{align*}
Here, $\xi_{d\sigma}$ refers to the energy of the dot orbital state $d$ for
spin $\sigma$, $\xi_{k\sigma}$ to the energy of lead state $k$ for spin
$\sigma$ and $t_{d\sigma}^{k}$ is an hoping matrix element (we assume that the
spin $\sigma$ is preserved upon tunneling like in
section~\ref{sec:non-interacting}). The index $k$ runs over the electronic
states of lead $L$ and $R$. Coulomb interactions are taken into account
through the term in $U=e^{2}/C_{\Sigma}$, with $n_{d\sigma}=c_{d\sigma
}^{\dagger}c_{d\sigma}$ and $C_{\Sigma}$ the total capacitance of the quantum
dot M. \textbf{\ }By construction of the model (see above), for $U=0$, each
orbital level $\xi_{d\sigma}$ corresponds to a resonant level $E_{d\sigma}%
^{c}$ of section~\ref{sec:non-interacting}, with $\xi_{d\downarrow}%
-\xi_{d\uparrow}=g\mu_{B}h_{SDIPS}^{c}$. We can therefore regard the effective
Zeeman splitting $h_{SDIPS}^{c}$ in model~(\ref{H}) as a generalization of the
SDIPS concept to the interacting case\cite{explanationWetzels}. The
specificity of this effective field, with respect to an ordinary external
field, is that it depends on the configuration $c$ of the ferromagnetic
electrodes. For instance, in the case of symmetric ferromagnetic contacts,
symmetry considerations lead to $h_{SDIPS}^{P}\neq0$ and $h_{SDIPS}^{AP}=0$.

In the following, we calculate the zero-bias conductance of the circuit
using~\cite{Meir2}
\begin{align}
\frac{h}{e^{2}}\frac{G^{c}}{2}  &  =\label{cond}\\
&  \sum\limits_{d,\sigma}\int_{-\infty}^{+\infty}d\omega\frac{\partial
f(\hbar\omega)}{\partial\hbar\omega}\frac{\Gamma_{d\sigma}^{L}(\hbar
\omega)\Gamma_{d\sigma}^{R}(\hbar\omega)}{\Gamma_{d\sigma}^{L}(\hbar
\omega)+\Gamma_{d\sigma}^{R}(\hbar\omega)}\operatorname{Im}[G_{d\sigma}%
(\omega)]\text{ .}\nonumber
\end{align}
The above equation involves the retarded Green's function $G_{d\sigma}%
(\omega)=\int_{-\infty}^{+\infty}\widetilde{G}_{d\sigma}(t)e^{i\omega t}%
dt$\ with $\widetilde{G}_{d\sigma}(t)=-i\theta(t)\left\langle \left\{
c_{d\sigma}(t),c_{d\sigma}^{\dagger}(0)\right\}  \right\rangle $. We also use
the Fermi distribution $f(\xi)=(1+\exp[\xi])^{-1}$\ and the tunnel transition
rates $\Gamma_{d\sigma}^{l}(\xi)=\sum\limits_{k}2\pi\left|  t_{d\sigma}%
^{k}\right|  ^{2}\delta(\xi=\xi_{k\sigma})$ with $l\in\{L,R\} $. Note that
$G_{d\sigma}$, $\xi_{d\sigma}$ and $\Gamma_{d\sigma}^{l}$ depend on the
configuration $c\in\{P,AP\}$ considered but for simplicity we omit the index
$c$ in those quantities. We want to study current transport in the limit
studied in Ref.~\onlinecite{Sahoo}, i.e. the width of conductance peaks
displayed by the circuit is determined not only by temperature but also by the
tunnel rates ($\Gamma_{d\sigma}^{L}+\Gamma_{d\sigma}^{R}$ $\sim2k_{B}T$). This
requires to go beyond the sequential tunneling description, i.e. to take into
account high-order quantum tunneling processes. For this purpose, we will
calculate $G_{d\sigma}$ using the equation of motion (E.O.M.)
technique\cite{Meir}, which is valid for temperatures larger than the Kondo
temperature $T_{K}$ of the system\cite{note SA}.

\section{Single level quantum dot}

\label{sec:single-level}

For simplicity, we first take into account a single orbital level $d$ of the
dot. We follow the lines of Ref.~\onlinecite{Meir}. The E.O.M. technique leads
to
\begin{equation}
\frac{G_{d\sigma}(\omega)}{\hbar}=\frac{1-\left\langle n_{d\overline{\sigma}%
}\right\rangle }{\hbar\omega-\xi_{d\sigma}-\Sigma_{\sigma}^{S}}+\frac
{\left\langle n_{d\overline{\sigma}}\right\rangle }{\hbar\omega-\xi_{d\sigma
}-\Sigma_{\sigma}^{D}} \label{Gtot}%
\end{equation}
with
\begin{equation}
\left\langle n_{d\sigma}\right\rangle =-\int_{-\infty}^{+\infty}\frac{d\omega
}{\pi}f(\hbar\omega)\operatorname{Im}[G_{\sigma}^{d}(\omega)]
\label{Article::eq:1}%
\end{equation}
the average occupation of orbital $d$ by electrons with spin $\sigma$. We
define
\begin{equation}
\Sigma_{\sigma}^{S}=\Sigma_{d\sigma}^{0}-U\Sigma_{d\sigma,d\overline{\sigma}%
}^{1,1}[\hbar\omega-\xi_{d\sigma}-U-\Sigma_{d\sigma}^{0}-\Sigma_{d\sigma
,d\overline{\sigma}}^{3,1}]^{-1}\,, \label{Article::eq:2}%
\end{equation}%
\begin{equation}
\Sigma_{\sigma}^{D}=U+\Sigma_{d\sigma}^{0}+U\Sigma_{d\sigma,d\overline{\sigma
}}^{2,1}[\hbar\omega-\xi_{d\sigma}-\Sigma_{\sigma}^{0}-\Sigma_{d\sigma
,d\overline{\sigma}}^{3,1}]^{-1}\,, \label{Article::eq:3}%
\end{equation}%
\begin{equation}
\Sigma_{d\sigma}^{0}=\sum\limits_{k}\left|  t_{d\sigma}^{k}\right|
^{2}\left(  \hbar\omega-\xi_{k\sigma}+i0^{+}\right)  ^{-1} \label{A}%
\end{equation}
and, for $i\in\{1,2,3\}$,
\begin{multline}
\Sigma_{d\sigma,d^{\prime}\sigma^{\prime}}^{i,n}=\sum\limits_{k}\frac{\mu
_{i}(\xi_{k\sigma^{\prime}})\left|  t_{k\sigma^{\prime}}^{d^{\prime}}\right|
^{2}}{\hbar\omega-\xi_{d\sigma}+\xi_{d^{\prime}\sigma^{\prime}}-\xi
_{k\sigma^{\prime}}+i0^{+}}\label{B}\\
\mbox{}+\sum_{k}\frac{\mu_{i}(\xi_{k\sigma^{\prime}})\left|  t_{k\sigma
^{\prime}}^{d^{\prime}}\right|  ^{2}}{\hbar\omega-\xi_{d\sigma}-\xi
_{d^{\prime}\sigma^{\prime}}-nU+\xi_{k\sigma^{\prime}}+i0^{+}}\,.
\end{multline}
Here, one has $\mu_{1}(\xi)=f(\xi)$, $\mu_{2}(\xi)=1-f(\xi)$ and $\mu_{3}=1$
(We anticipate on the next paragraphs by defining $\Sigma_{d\sigma,d^{\prime
}\sigma^{\prime}}^{i,n}$ for $n\in\mathbb{N}$ and an arbitrary dot state
$d^{\prime}\sigma^{\prime}\neq d\sigma$, but only $n=1$ and $d^{\prime}%
\sigma^{\prime}=d\overline{\sigma}$ are needed for the present one-orbital
case). We assume that the coupling to the leads is energy independent (broad
band approximation), which gives e.g. $\Sigma_{d\sigma}^{0}=-i(\Gamma
_{d\sigma}^{L}+\Gamma_{d\sigma}^{R})/2$. The term $\Sigma_{d\sigma}^{0}$,
which is due to the tunneling of electrons with spin $\sigma$, already
occurred in the non-interacting case\cite{RelModel}. \begin{figure}[ptb]
\centering\includegraphics[width=1\linewidth]{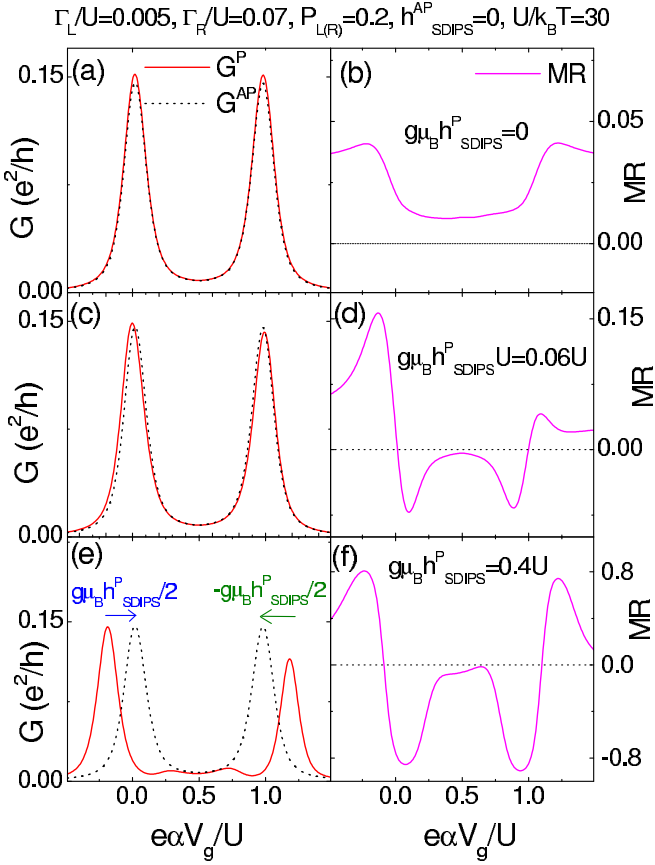}\caption{Panels a, c and
e: Conductance $G^{P}$ in the parallel configuration (red full lines) and
conductance $G^{AP}$ in the antiparallel configuration (black dotted lines) as
a function of the gate voltage $V_{g}$, for the circuit shown in Fig.
\ref{Figure1}, with M a 1-orbital quantum dot. We have used $\Gamma
_{L}=0.005U$, $\Gamma_{R}=0.07U$, $P_{L(R)}=0.2$, $U/k_{B}T=30$ and
$h_{SDIPS}^{AP}=0$. Panels b, d, and f: Magnetoresistance $\mathrm{MR}%
=(G^{P}-G^{AP})/(G^{P}+G^{AP})$ (pink curves) corresponding to the left
conductance plots. The results are shown for $g\mu_{B}h_{SDIPS}^{P}=0$ (panels
a and b), $g\mu_{B}h_{SDIPS}^{P}=0.06U$ (panels c and d) and $g\mu
_{B}h_{SDIPS}^{P}=0.4U$ (panels e and f).}%
\label{Figure2}%
\end{figure}\begin{figure}[ptbptb]
\centering\includegraphics[width=1.\linewidth]{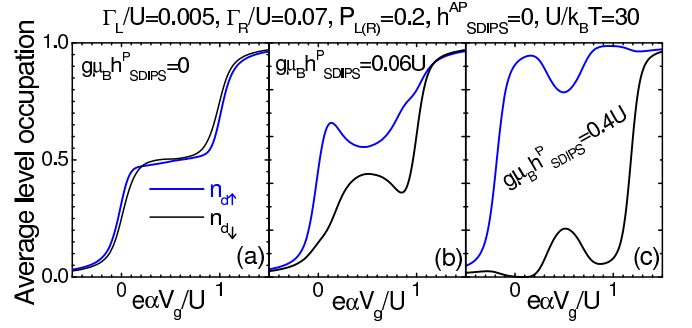}\caption{Average
occupations $\left\langle n_{d\uparrow}\right\rangle $ (blue lines) and
$\left\langle n_{d\downarrow}\right\rangle $ (black lines) of level $d$ by
spins $\uparrow$ and $\downarrow$\ as a function of $V_{g}$, for the 1-orbital
quantum dot circuit of Fig. \ref{Figure1}. The results are shown for the same
parameters as in Fig. \ref{Figure2} and lead polarizations in the parallel
configuration ($c=P$), with $h_{SDIPS}^{P}=0$ (panel a), $g\mu_{B}%
h_{SDIPS}^{P}=0.06U$, (panel b) and $g\mu_{BSDIPS}^{P}=0.4U$ (panel c). For
$h_{SDIPS}^{P}=0$, $\left\langle n_{d\uparrow}\right\rangle $ and
$\left\langle n_{d\downarrow}\right\rangle $ remain very close, simply showing
two steps corresponding to the two conductances peaks visible in Fig.
\ref{Figure2}-a. In the case of a finite $h_{SDIPS}^{P}$, $\left\langle
n_{d\uparrow}\right\rangle $ rises more strongly than $\left\langle
n_{d\downarrow}\right\rangle $ at the first conductance peak, revealing that
current transport is due in majority to $\uparrow$ spins for this first peak.
Then, $\left\langle n_{d\uparrow}\right\rangle $ and $\left\langle
n_{d\downarrow}\right\rangle $ become closer when both $\xi_{d\uparrow}$ and
$\xi_{d\downarrow}$ are below the Fermi level. At the second conductance peak
$\left\langle n_{d\downarrow}\right\rangle $ rises more strongly than
$\left\langle n_{d\uparrow}\right\rangle $ because current transport is now
dominated by down spins. The asymmetry between the behaviors of spins
$\uparrow$ and $\downarrow$ increases with $h_{SDIPS}^{P}$ (from left to right
panels). }%
\label{Figure1N}%
\end{figure}In the interacting case, $G_{d\sigma}(\omega)$ also involves
$\Sigma_{d\sigma,d^{\prime}\sigma^{\prime}}^{i,n}$ terms related to the
tunneling of electrons with spin $\overline{\sigma}$. The average occupation
$\left\langle n_{d\sigma}\right\rangle $ can be calculated from Eqs.
(\ref{Gtot}) and (\ref{Article::eq:1}) as
\begin{equation}
\left\langle n_{d\sigma}\right\rangle =\frac{\left\langle n_{S\sigma
}\right\rangle (1-\left\langle n_{S\overline{\sigma}}\right\rangle
)+\left\langle n_{S\overline{\sigma}}\right\rangle \left\langle n_{D\sigma
}\right\rangle }{1-(\left\langle n_{D\sigma}\right\rangle -\left\langle
n_{S\sigma}\right\rangle )(\left\langle n_{D\overline{\sigma}}\right\rangle
-\left\langle n_{S\overline{\sigma}}\right\rangle )} \label{Article::eq:4}%
\end{equation}
with, for $j\in\{S,D\}$,
\begin{equation}
\left\langle n_{j\sigma}\right\rangle =-\frac{1}{\pi}\operatorname{Im}%
\int_{-\infty}^{+\infty}\hbar{d\omega}\;\frac{f(\hbar\omega)}{\hbar\omega
-\xi_{d\sigma}-\Sigma_{\sigma}^{j}(\omega)} \label{Article::eq:5}%
\end{equation}

Figure \ref{Figure2} shows the conductance $G^{c}$ in configuration
$c\in\{P,AP\}$ (panels a, c and e) and the magnetoresistance $\mathrm{MR}%
=(G^{P}-G^{AP})/(G^{P}+G^{AP})$ (panels b, d and f) calculated for different
values of $h_{SDIPS}^{c}$, using $\Gamma_{d\uparrow\lbrack\downarrow]}%
^{l}=\Gamma_{l}(1\pm P_{l})$ for $l\in\{L,R\}$. We have used parameters
consistent with Ref.~\onlinecite{Sahoo}, i.e. $U/k_{B}T=30$, and tunnel rates
$\Gamma_{L(R)}$ leading to the proper width and height for the conductance
peaks. We have also used relatively low values for $P_{L(R)}$ because usual
ferromagnetic contact materials are not fully polarized \cite{Soulen}. The
conductance peak corresponding to level $d$ is split due to Coulomb
interactions (see Eq. (\ref{Gtot}), Figs. \ref{Figure2}-a, \ref{Figure2}-c and
\ref{Figure2}-e). At low temperatures $T<T_{K}$, Kondo effect is expected in
the valley between the two resulting peaks. We have checked that the
hypothesis $T>T_{K}$ and hence the E.O.M. technique are valid for the
parameters of Fig. \ref{Figure2} (see Refs. \onlinecite{TKferro, Aleiner}).
For $h_{SDIPS}^{c}=0$, we already note a strong qualitative difference with
the non-interacting case: although the two conductance peaks displayed by
$G^{P}(V_{g})$ are very similar, the $\mathrm{MR}$ variations corresponding to
these two peaks have different shapes\cite{Weymann}. More precisely, for the
low values of polarization considered here, $\mathrm{MR}(V_{g})$ is
approximately mirror symmetric from one conductance peak to the other. Note
that in Fig. \ref{Figure2}, we have used specific parameters such that
$\mathrm{MR}$ remains positive for any value of $V_{g}$ when there is no
SDIPS. Nevertheless, it is possible to have $MR<0$ for $h_{SDIPS}%
^{P}=h_{SDIPS}^{AP}=0$, for instance by increasing $P_{L(R)}$ (not shown).

We now address the effect of a finite effective field $h_{SDIPS}^{c}$. This
field produces a shift of the conductance peaks from the $P$ to the $AP$
configurations. For instance, in Fig. 2-c and Fig. 2-e, plotted for
$h_{SDIPS}^{P}\neq0$ and $h_{SDIPS}^{AP}=0$, the left [right] conductance peak
is shifted to the right [left] from $P$ to $AP$ because it mainly comes from
the transport of up [down] spins\ in the $P$\ case (this can be seen from the
average occupations of the levels versus $V_{g}$ in Fig. \ref{Figure1N}). As a
consequence, in Fig. \ref{Figure2}, $\mathrm{MR}$ becomes negative for certain
values of gate voltage. The effective field $h_{SDIPS}^{c}$ thus enhances
negative $\mathrm{MR}$ effects. If $h_{SDIPS}^{c}$ is strong enough, it can
even produce a giant $\mathrm{MR}$ effect with its sign tunable with $V_{g}$
(Fig.~\ref{Figure2}-f). Moreover, because of the opposite shifts of the two
consecutive conductance peaks for $c=P$ with respect to those for $c=AP$, the
positive-and-then-negative profile of MR corresponding to one conductance peak
is generally followed by the negative-and-then-positive profile near the next
conductance peak (approximately mirror symmetric). Note that a sign change
$h_{SDIPS}^{P}\rightarrow-h_{SDIPS}^{P}$ will not modify this behavior for the
low values of polarizations considered here, because the spin with
lower[higher] orbital energy will dominate in the left [right] peak of
$G^{P}(V_{g})$.

We now compare the results of this section with the experimental data of Ref.
\onlinecite{Sahoo}. Like many Coulomb blockade devices, the circuit studied in
this experiment suffered from low frequency $V_{g}$-noise, which can be
attributed to charge fluctuators located in the vicinity of the device. A
strong gate voltage offset jump occurred at $V_{g}=4.331~$\textrm{V}, and the
data before and after this jump do not necessarily correspond to the filling
of consecutive levels. Therefore, we will focus on the data taken for
$V_{g}>4.331~$\textrm{V}$,$ shown \cite{PC} with black squares in Fig.
\ref{Figure4}. These data display almost 2 regular $\mathrm{MR}(V_{g})$
oscillations, which cannot be understood with the 1-orbital model. Indeed, as
explained above, in this model, the two conductance peaks of $G^{c}(V_{g}) $
are shifted in opposite directions by $h_{SDIPS}^{c}$. As a consequence, the
$\mathrm{MR}(V_{g})$ variations corresponding to these two peaks cannot be
similar for parameters consistent with the experiment. We have shown here
curves for $h_{SDIPS}^{AP}=0$, but a finite $h_{SDIPS}^{AP}$ would not modify
this result. Using values of $P_{L(R)}$ larger than in Fig.~\ref{Figure2}
would not help either.

For simplicity, we have considered in this section the one-orbital case. In
reality, there are more than one orbital levels on a quantum dot. As long as
these levels are sufficiently well separated from each other (roughly, by an
orbital energy difference larger than the Hund-rule exchange energy), the two
conductance peaks associated to a given level will occur consecutively in
$G^{c}(V_{g})$ and will thus be described qualitatively like above\cite{Meir}.
In particular, the two peaks will be shifted in opposite directions by
$h_{SDIPS}^{c}\neq0$; the first peak to lower values of $V_{g}$ and the second
peak to higher values. Therefore, this limit should not allow to obtain two
consecutive conductance peaks with analogue $MR(V_{g})$\ patterns. On the
contrary, if two (or more) levels are nearly degenerate, it is possible that
the orbital levels of the quantum dot are not filled one by one while
increasing $V_{g}$. Therefore, consecutive conductance peaks may exhibit a
qualitatively different behavior compared with the one-orbital case. To
examine this effect, we will consider in next section the extreme case of a
quantum dot with a twofold orbital degeneracy. We will see that the
discrepancy between the theory and the data can be resolved by using this model.

Before concluding this section, we make a remark on another possible
contribution to the spin splitting of the conductance peaks. Even though so
far we have mainly considered the contribution from the SDIPS, in principle,
virtual particle exchange processes with the spin-polarized leads can also
renormalize the energy levels through the $\Sigma_{d\sigma,d^{\prime}%
\sigma^{\prime}}^{i,n}$ terms of Eq.~(\ref{B}) \cite{pointout,CorrExch}.
Indeed, the $\Sigma_{d\sigma,d^{\prime}\sigma^{\prime}}^{i,n}$ terms are not
negligible in general. For example, in Fig. \ref{Figure2}-a, they globally
shift the position of the conductance peaks in $G^{P}(V_{g})$ by about $3.2\%
$\ of $U/e\alpha$. Nevertheless,\ for the low values of polarizations
$P_{L(R)}$\ and the temperatures used here, the level spin-splitting produced
by the $\Sigma_{d\sigma,d^{\prime}\sigma^{\prime}}^{i,n}$ terms is much weaker
than this global shift and cannot compete with the finite values of
$h_{SDIPS}^{c}$ considered in this article.

\section{Quantum dot with a doubly-degenerate level}

\label{sec:two-level}

In order to improve the understanding of Ref. \onlinecite{Sahoo}, we\ now take
into account the $K$-$K^{^{\prime}}$ orbital degeneracy commonly
observed\cite{Liang, Jarillo, Sapmaz, Moriyama, Babic, Babic2} in SWNTs, by
considering a two-orbitals model i.e. hamiltonian (\ref{H}) with
$d\in\{K,K^{^{\prime}}\}$ and $\xi_{K^{\prime}\sigma}=\xi_{K\sigma}$.
Interestingly, SU(4) Kondo effect involving the orbital and spin degrees of
freedom was observed in SWNTs with the $K$-$K^{^{\prime}}$ degeneracy
\cite{SU4exp,SU4th}. This suggests that, in this system, the orbital quantum
number is conserved during higher order tunnel events, probably because the
electrons of the nanotube quantum dot are coupled to the nanotube section
underneath the contacts, where they dwell for some time before moving into the
metal. For simplicity, we will also assume such a situation here and disregard
high-order quantum processes which couple the $K$ and $K^{^{\prime}}$
orbitals~\cite{IfNot}. In order to calculate the conductance of the system
from Eq. (\ref{cond}), one needs to calculate the retarded Green's function
$G_{s}(\omega)$\ for $s\in\{\{K\uparrow\},\{K\downarrow\},\{K^{^{\prime}%
}\uparrow\},\{K^{^{\prime}}\downarrow\}\}$. For this purpose, we again use the
EOM technique. Since it is not possible to obtain a simple analytical
expression for $G_{s}(\omega)$\ in the two-orbitals case, we show below the
system of equations of motion calculated by neglecting electronic correlations
between the dot and the leads ($T>T_{K}$). Using $s$, $s_{1}$, $s_{2}$\ and
$s_{3}$\ to denote four different dot states in the ensemble $\{\{K\uparrow
\},\{K\downarrow\},\{K^{^{\prime}}\uparrow\},\{K^{^{\prime}}\downarrow\}\}$,
we obtain \begin{figure}[ptb]
\centering\includegraphics[width=1\linewidth]{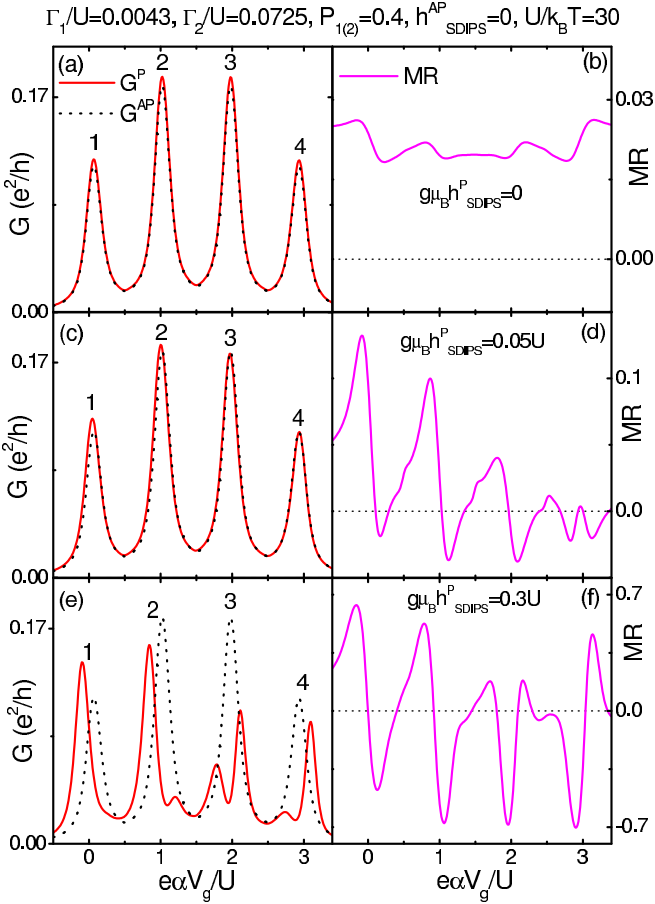}\caption{Panels a, c and
e: Conductance $G^{P}$ in the parallel configuration (red full lines) and
conductance $G^{AP}$ in the antiparallel configuration (black dotted lines),
for the circuit of Fig. \ref{Figure1}, with M a two-orbitals quantum dot. We
have used identical tunnel rates to the two orbitals, i.e. $\Gamma
_{L}=0.0043U$, $\Gamma_{R}=0.0725U$, and $P_{L(R)}=0.4$. We have also used
$U/k_{B}T=30$ and $h_{SDIPS}^{AP}=0$. Panels b, d and f: Magnetoresistance
$\mathrm{MR}$ (pink full lines) corresponding to the left conductance plots.
The results are shown for $g\mu_{B}h_{SDIPS}^{P}=0$ (panels a and b),
$g\mu_{B}h_{SDIPS}^{P}=0.05U$ (panels c and d) and $g\mu_{B}h_{SDIPS}%
^{P}=0.3U$ (panels e and f). }%
\label{Figure3}%
\end{figure}%
\begin{equation}
G_{s}=\left(  \hbar\omega-\xi_{s}-\Sigma_{s}^{0}\right)  ^{-1}\left\{
\hbar+U\left(  D_{s}^{s_{1}}+D_{s}^{s_{2}}+D_{s}^{s_{3}}\right)  \right\}
\label{A1}%
\end{equation}%
\begin{align}
D_{s}^{s_{1}}  &  =\left(  \hbar\omega-\xi_{s}-U-\Sigma_{s}^{0}-\Sigma
_{s,s_{1}}^{3,1}\right)  ^{-1}\left\{  \hbar\left\langle n_{s_{1}%
}\right\rangle -\Sigma_{s,s_{1}}^{1,1}G_{s}\right. \nonumber\\
&  +\left(  U-\chi_{s,s_{1}}^{3,1}\right)  \left(  D_{s}^{s_{1},s_{2}}%
+D_{s}^{s_{1},s_{3}}\right)  +\chi_{s,s_{1}}^{1,1}\left(  D_{s}^{s_{2}}%
+D_{s}^{s_{3}}\right) \nonumber\\
&  +\left.  \left(  \chi_{s,s_{1}}^{1,3}-\chi_{s,s_{1}}^{1,1}\right)
D_{s}^{s_{2},s_{3}}+\left(  \chi_{s,s_{1}}^{3,1}-\chi_{s,s_{1}}^{3,3}\right)
D_{s}^{s_{1},s_{2},s_{3}}\right\}  \label{A2}%
\end{align}%
\begin{align}
D_{s}^{s_{1},s_{2}}  &  =\left(  \hbar\omega-\xi_{s}-2U-\Sigma_{s}^{0}%
-\Sigma_{s,s_{1}}^{3,3}-\Sigma_{s,s_{2}}^{3,3}\right)  ^{-1}\nonumber\\
&  \left\{  \hbar\left\langle n_{s_{1}}n_{s_{2}}\right\rangle -\Sigma
_{s,s_{1}}^{1,3}D_{s}^{s_{2}}-\Sigma_{s,s_{2}}^{1,3}D_{s}^{s_{1}}%
+\chi_{s,s_{1}}^{1,3}D_{d\sigma}^{s_{2},s_{3}}\right. \nonumber\\
&  \left.  +\chi_{s,s_{2}}^{1,3}D_{s}^{s_{1},s_{3}}+\left(  U-\chi_{s,s_{1}%
}^{3,3}-\chi_{s,s_{2}}^{3,3}\right)  D_{s}^{s_{1},s_{2},s_{3}}\right\}
\label{A3}%
\end{align}%
\begin{align}
D_{s}^{s_{1},s_{2},s_{3}}  &  =\left(  \hbar\omega-\xi_{s}-3U-\Sigma_{s}%
^{0}-\Sigma_{s,s_{1}}^{3,5}-\Sigma_{s,s_{2}}^{3,5}-\Sigma_{s,s_{3}}%
^{3,5}\right)  ^{-1}\nonumber\\
&  \left\{  \hbar\left\langle n_{s_{1}}n_{s_{2}}n_{s_{3}}\right\rangle \right.
\nonumber\\
&  \left.  -\Sigma_{s,s_{1}}^{1,5}D_{s}^{s_{2},s_{3}}-\Sigma_{s,s_{2}}%
^{1,5}D_{s}^{s_{3},s_{1}}-\Sigma_{s,s_{3}}^{1,5}D_{s}^{s_{1},s_{2}}\right\}
\label{A4}%
\end{align}
Due to interaction $U$, the Green's function $G_{s}=G_{s}(\omega)$\ is coupled
to other Green's functions $D_{s}^{s_{1},...,s_{3}}(\omega)=\int_{-\infty
}^{+\infty}\widetilde{D}_{s}^{s_{1},...,s_{3}}(\omega)e^{i\omega t}dt$\ with
$\widetilde{D}_{s}^{s_{1},..,s_{_{3}}}(t)=-i\theta(t)\langle\{n_{s_{1}%
}(t)..n_{s_{3}}(t)c_{s}(t),c_{s}^{\dagger}\}\rangle$\ and $n_{s_{i}%
}(t)=c_{s_{i}}^{\dagger}(t)c_{s_{i}}(t)$\ for $i\in\{1,3\}$. This means that
the dynamics of electrons in state $s$\ is modified by the presence of other
electrons on the dot [In the one orbital case, $G_{d\sigma} $\ was coupled to
$D_{d\sigma}^{d\overline{\sigma}}$\ only, which lead to simple expression
(\ref{Gtot})]. The term $\Sigma_{s}^{0}=-i(\Gamma_{s}^{L}+\Gamma_{s}^{R}%
)/2$\ is the tunneling self energy for a non-interacting quantum dot, already
introduced in previous section. The equations of motion also involve terms
$\Sigma_{d\sigma,d^{\prime}\sigma^{\prime}}^{i,n}$, which are defined by
Eq.(\ref{B}), and terms defined by
\begin{multline}
\chi_{d\sigma,d^{\prime}\sigma^{\prime}}^{i,n}=\sum\limits_{k}\left(
\frac{\mu_{i}(\xi_{k\sigma^{\prime}})\left\vert t_{k\sigma^{\prime}%
}^{d^{\prime}}\right\vert ^{2}}{\hbar\omega-\xi_{d\sigma}-\xi_{d^{\prime
}\sigma^{\prime}}+\xi_{k\sigma^{\prime}}-nU+i0^{+}}\right.
\label{Article::eq:6}\\
\left.  -\frac{\mu_{i}(\xi_{k\sigma^{\prime}})\left\vert t_{k\sigma^{\prime}%
}^{d^{\prime}}\right\vert ^{2}}{\hbar\omega-\xi_{d\sigma}-\xi_{d^{\prime
}\sigma^{\prime}}+\xi_{k\sigma^{\prime}}-(n+2)U+i0^{+}}\right)
\end{multline}
These terms take into account the tunneling of electrons between the leads and
a dot state $d^{\prime}\sigma^{\prime}$\ different from $d\sigma$. The average
level occupations occuring in Eqs.~(\ref{A1})--(\ref{A4}) are given by
$\left\langle n_{s_{1}},..,n_{s_{n}}\right\rangle =-\int_{-\infty}^{+\infty
}\frac{d\omega}{\pi}f(\hbar\omega)\operatorname{Im}[D_{n_{s_{n}}}%
^{s_{1},..,s_{n-1}}(\omega)]$. The Green's functions and the level occupations
can be calculated numerically from the above equations. Up to now, the E.O.M.
technique for multilevel systems had been implemented only by neglecting
$\Sigma_{d\sigma,d^{\prime}\sigma^{\prime}}^{i,n}$ and $\chi_{d\sigma
,d^{\prime}\sigma^{\prime}}^{i,n}$ terms \cite{multi}. Like in the one-orbital
case, these terms are not negligible in the context of our study.
\begin{figure}[ptbptb]
\centering\includegraphics[width=1.\linewidth]{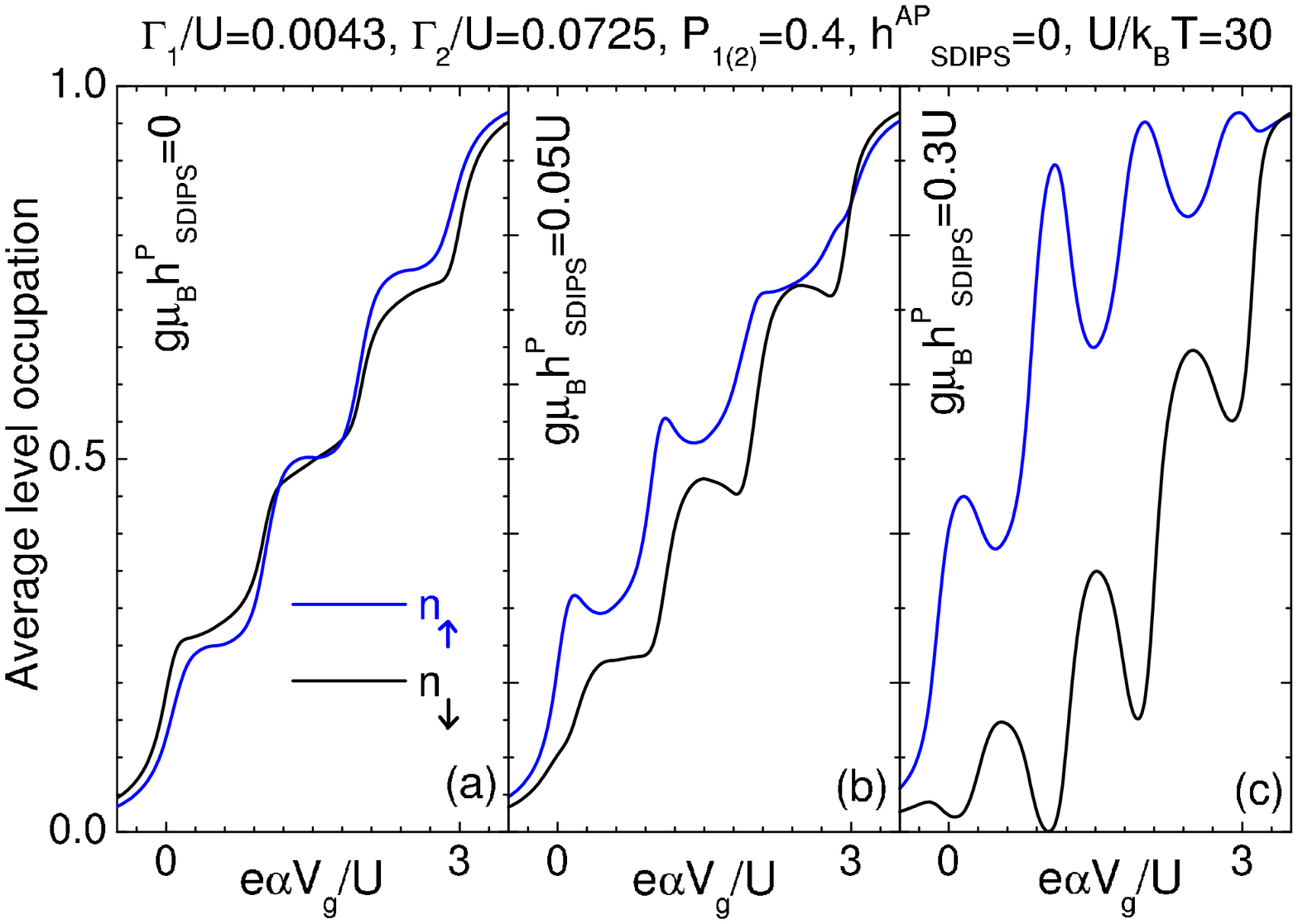}\caption{Average
occupations $\left\langle n_{\uparrow}\right\rangle =\left\langle
n_{K\uparrow}\right\rangle =\left\langle n_{K^{\prime}\uparrow}\right\rangle $
(blue lines) and $\left\langle n_{\downarrow}\right\rangle =\left\langle
n_{K\downarrow}\right\rangle =\left\langle n_{K^{\prime}\downarrow
}\right\rangle $ (black lines) of levels $K$ and $K^{\prime}$ by spins
$\uparrow$ and $\downarrow$\ as a function of $V_{g}$, for a two-orbitals
quantum dot circuit with the same parameters as in Fig. \ref{Figure3}. The
results are shown for lead polarizations in the parallel configuration
($c=P$), with $h_{SDIPS}^{P}=0$ (panel a), $g\mu_{B}h_{SDIPS}^{P}=0.05U$
(panel b), and $g\mu_{B}h_{SDIPS}^{P}=0.3U$ (panel c). For $h_{SDIPS}^{P}=0$,
$\left\langle n_{\uparrow}\right\rangle $ and $\left\langle n_{\downarrow
}\right\rangle $ remains very close, showing four steps corresponding to the
four conductances peaks visible in Fig. \ref{Figure3}-a. In the case of a
finite $h_{SDIPS}^{P}$, $\left\langle n_{\uparrow}\right\rangle $ rises more
strongly than $\left\langle n_{\downarrow}\right\rangle $ for the two first
conductance peaks, which shows that current transport is due in majority to up
spins for these two peaks. The opposite situation occurs for the two last
conductance peaks. The asymmetry between the behaviors of spins $\uparrow$ and
$\downarrow$ increases with $h_{SDIPS}^{P}$ (from left to right panels).}%
\label{Figure2N}%
\end{figure}

Figure~\ref{Figure3} shows the conductance (panels a, c and e) and
$\mathrm{MR}$ curves (panels b, d and f) calculated from Eqs. (\ref{cond}) and
(\ref{A1}-\ref{A4}), for different values of $h_{SDIPS}^{c}$. For simplicity,
we have assumed that the coupling to the leads is identical for the two
orbitals, i.e. $\Gamma_{K\uparrow\lbrack\downarrow]}^{l}=\Gamma_{K^{^{\prime}%
}\uparrow\lbrack\downarrow]}^{l}=\Gamma_{l}(1\pm P_{l})$ for $l\in\{L,R\}$. We
have again used parameters consistent with Ref. \onlinecite{Sahoo} i.e.
$U/k_{B}T=30$, relatively low polarizations $\left\vert P_{L(R)}\right\vert
=0.4$ and values of $\Gamma_{L(R)}$ leading to the proper width and height for
the conductance peaks. We have checked that these parameters are compatible
with the hypothesis $T>T_{K}$, with $T_{K}$ the Kondo temperature associated
to the SU(4) Kondo effect expected in this system \cite{SU4th}. In most cases,
the curves $G^{c}(V_{g})$ show 4 resonances, the first two associated with a
single occupation of $K$ and $K^{\prime}$, and the other two to double
occupation (see e.g. Fig. \ref{Figure3}-a). For $h_{SDIPS}^{P}=h_{SDIPS}%
^{AP}=0$ and the parameters used here, $\mathrm{MR}$ remains positive for any
value of $V_{g}$ (Fig. \ref{Figure3}-b). Like in the 1-orbital case, a finite
$h_{SDIPS}^{c}$ makes easier negative $\mathrm{MR}$ effects and can even lead
to a giant $\mathrm{MR}$ effect with a sign tunable with $V_{g}$ (Figs 4-d and
4-f). Importantly, the effect of $h_{SDIPS}^{c}$ depends on the occupation of
the dot. For instance, in Fig. \ref{Figure3}-e plotted for $g\mu_{B}%
h_{SDIPS}^{P} $ larger than the linewidth of the conductance peaks, the first
two conductance peaks of $G^{P}$ (peaks 1 and 2) are strongly shifted to the
left by $h_{SDIPS}^{P}$ because they are due in majority to up spins, as can
be seen from the average occupation of the levels in Fig. \ref{Figure2N}-c.
This allows to get a $\mathrm{MR}$ pattern approximately similar for these two
peaks, i.e. a transition from positive to negative values of $\mathrm{MR} $
(Fig. \ref{Figure3}-f). On the contrary, peak 4 corresponds to a transition
from negative to positive values of $\mathrm{MR}$ because the associated
conductance peak is due in majority to down spins. In Fig. \ref{Figure3}-e,
the shape of the $\mathrm{MR}(V_{g})$ pattern associated to peak 3 is more
particular (positive/negative/positive) because, for the values of parameters
considered here, Coulomb blockade does not entirely suppress the up spins
contribution in peak 3, which is therefore spin-split \cite{MoreSym}.
Remarkably, this allows to obtain, at the left of Fig. \ref{Figure3}-f, three
positive $\mathrm{MR}$ maxima which differ in amplitude but have rather
similar shapes. In the case of $g\mu_{B}h_{SDIPS}^{P}$ finite but smaller than
the linewidth of the conductance peaks (Fig. \ref{Figure3}-c), the amplitude
of the $\mathrm{MR}$ signal is much smaller than in the previous case but its
shape remains comparable. \begin{figure}[ptb]
\centering\includegraphics[width=1\linewidth]{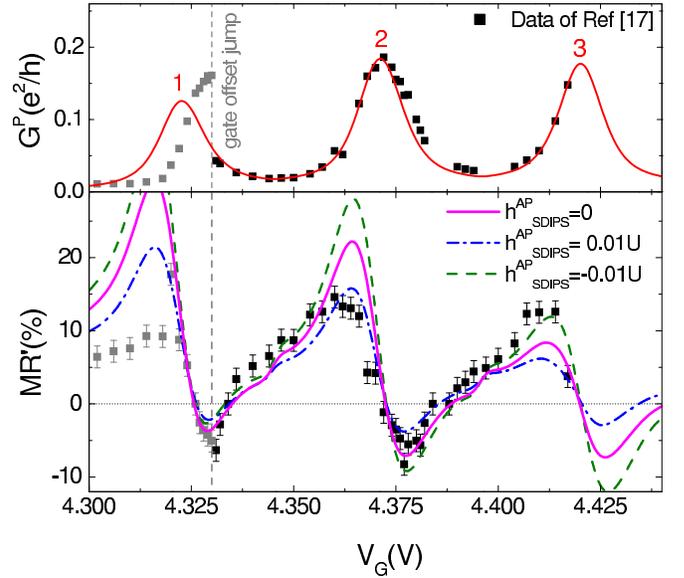}\caption{Comparison
between the data of Ref. \onlinecite{Sahoo} (squares) and the two-orbitals
theory. We show the conductance $G^{P}$ in the parallel configuration (top
panel) and the corresponding magnetoresistance $\mathrm{MR}^{\prime}%
=(G^{P}-G^{AP})/G^{AP}$ (bottom panel). The theory is shown for parameters
consistent with the experiment i.e. $U=5~\mathrm{meV}$, $U/k_{B}T=30$ and
$\alpha=0.0986$. We also use relatively low values of polarization
$P_{L(R)}=0.4$ because usual ferromagnetic contact materials are not fully
polarized \cite{Soulen}. Assuming identical tunnel couplings for the two
orbitals, the values of tunnels rates $\Gamma_{L}=0.0043U$ and $\Gamma
_{R}=0.0725U$ are imposed by the width and height of the conductance peaks.
Then, $h_{SDIPS}^{P[AP]}$ are the only free fitting parameters which remain
for interpreting the $\mathrm{MR}$ curve. We have assumed $g\mu_{B}%
h_{SDIPS}^{P}=0.05U$\ for all the theoretical curves shown in this Figure. We
have plot the $MR$\ curves of bottom panel for $h_{SDIPS}^{AP}=0$\ (pink
curve, corresponding to Fig. \ref{Figure3}-d), $h_{SDIPS}^{AP}=-0.01U$\ (green
dashed curve) and $h_{SDIPS}^{AP}=0.01U$\ (blue dot-dashed curve). Note that
in this Figure, we show $MR^{\prime}$\ instead of $MR=(G^{P}-G^{AP}%
)/(G^{P}+G^{AP}$\ $)$\ in order to be consistent with Ref.~ \onlinecite
{Sahoo}. A strong gate voltage offset jump occurred at $V_{g}=4.331~$V,
therefore, we show the data at the left/right of this jump with grey/black
symbols.}%
\label{Figure4}%
\end{figure}

We now reconsider the experimental data of Ref.~\onlinecite{Sahoo}. Even the
two-orbital model cannot not provide a reasonable fit to the data if we assume
$h_{SDIPS}^{P}=0$ and $h_{SDIPS}^{AP}=0$. In contrast, the two-orbital model
exhibits a good agreement with the experimental data for $h_{SDIPS}^{P}%
=0.05U$, $h_{SDIPS}^{AP}=0$, $\Gamma_{L}/U=0.0043$, $\Gamma_{R}/U=0.0725$,
$\left\vert P_{L(R)}\right\vert =0.4$, and parameters $U=5~meV$, $U/k_{B}%
T=30$, and $\alpha=0.0986$ given by the experiment (see Fig.~\ref{Figure4},
red and pink full curves).

We now discuss the value of $h_{SDIPS}^{P}=0.05U$ found for the above fit.
This corresponds to a magnetic field of about $2~\mathrm{T}$, which is too
strong to be attributed to stray fields from the ferromagnetic electrodes (see
e.g. Ref. \onlinecite{stray}). This is in favor of generalizing the SDIPS
concept to SWNTs quantum dots circuits, i.e. considering that the energy
levels of the dot are spin-split because the confinement potential created by
the ferromagnetic electrodes is spin-dependent. For comparison, we have
estimated $h_{SDIPS}^{P}$ in the non-interacting theory\cite{wire2005}, using
realistic parameters i.e leads with a Fermi energy $10$~\textrm{eV} and a
density of states polarized by $40\%$, and a nanotube with Fermi wavevector
$8.5$~$10^{9}\mathrm{m}^{-1}$, Fermi velocity\cite{bockrath} $v_{F}^{M}%
=8$~$10^{5}\mathrm{m.s}^{-1}$, length $L=500$~\textrm{nm} like in Ref.
\onlinecite{Sahoo}, and density of states $N_{F}^{M}=2L/\pi\hbar v_{F}^{M}$.
We have modeled the interfaces between the nanotube and the leads with Dirac
potential barriers\cite{note}, with a height which is spin-polarized by $40\%$
and an average value which corresponds to\cite{RelModel} $\Gamma_{L(R)}=$
$T_{L(R)}/2\pi N_{F}^{L(R)}\sim60$~$\mathrm{\mu}$\textrm{eV }(For comparison
the fitting parameters used in Fig.~\ref{Figure4} correspond to $\Gamma
_{L}=21$~$\mathrm{\mu}$\textrm{eV} and $\Gamma_{R}=362$~$\mathrm{\mu}%
$\textrm{eV}). We obtain $h_{SDIPS}^{P}\sim1.3~\mathrm{T}$, which is
consistent with the above analysis.

In the above discussion, we have assumed $h_{SDIPS}^{AP}=0$ for simplicity.
The height of the conductance peaks in the data imposes to use a strong
assymetry $\Gamma_{R}/\Gamma_{L}\sim17$\ between the left and right tunnel
rates. Thus, the two tunnel barriers are not symmetric, and there is no
fundamental reason to assume $h_{SDIPS}^{AP}=0$. Figure~\ref{Figure4} shows
examples of $MR$\ curves plotted for a finite $h_{SDIPS}^{AP}$. Using
$h_{SDIPS}^{AP}=0.01U$\ (green dashed curve) enhances the fit of the $MR$\ at
peak 2 whereas $h_{SDIPS}^{AP}=-0.01U$\ (blue dot-dashed curve) enhances the
fit of the $MR$\ at peak 3. Interestingly, with the non-interacting model,
assuming the most simple situation in which $\varphi_{l}^{P,\uparrow}%
-\varphi_{l}^{P,\downarrow}$\ has the same sign for the two leads, one finds
$\left|  h_{SDIPS}^{AP}\right|  <\left|  h_{SDIPS}^{P}\right|  $,\ which is in
agreement with the values used here. The fact that the best fit for the
$MR$\ patterns at peaks 2 and 3 correspond to different values of
$h_{SDIPS}^{AP} $\ might be due to a gate dependence of the SDIPS. This is
indeed possible since the potential profile of the interfaces between the wire
and the leads can vary with $V_{g}$.

We now comment briefly on the data taken for $V_{g}<4.331~$V. It is not sure
that the data shown before and after $V_{g}=4.331~$V correspond to the filling
of consecutive levels because of the gate voltage jump which occured at this
value of $V_{g}$. Nevertheless, the shape of the MR curve corresponding to
$V_{g}<4.331~$V is rather consistent with the theory shown in Fig. 4. This
suggests that these data really correspond to peak 1. At this stage, it is
important to point out that other orbital levels not taken into account in our
calculation should slightly modify the conductance peaks 1 and 4. The
discrepancy between the theory and the data for $V_{g}<4.331~$V could be
explained by the effect of the other orbitals. For the data at $V_{g}%
>4.331~$V, our fit is more quantitative since we have used peaks 2 and 3 of
the theory.

In principle, the modelisation of the orbital levels in SWNTs can be refined
by taking into account an exchange energy $J$\ which favors spin alignment, an
excess Coulomb energy $\delta U$\ related to the double occupation of the same
orbital, and a subband mismatch $\delta=\xi_{K^{\prime}\sigma}-\xi_{K\sigma
}\neq0$\ (see Ref. \onlinecite{Oreg}). In practice, $\delta U$\ is rather
small but $J$\ and $\delta$\ can be of the same order as $U$. Two different
regimes of parameters can occur in practice. If $\delta>J+\delta U+\left|
g\mu_{B}h_{SDIPS}^{c}\right|  $, two electrons with opposite spins will fill
consecutively the same energy level while $V_{g}$ increases (see Refs.
\onlinecite{Sapmaz,Moriyama}), and the behavior of the device should thus be
analogue to the non-degenerate multi-orbital case evoked at the end of section
\ref{sec:single-level}.\ Nevertheless, if $\delta<J+\delta U+\left|  g\mu
_{B}h_{SDIPS}^{c}\right|  $, peaks 1 and 2 [3 and 4] will correspond in
majority to the same spin direction, as observed experimentally by
\onlinecite{Liang, Jarillo}. In this case, the effect of $h_{SDIPS}^{c}%
$\ should be qualitatively the same as described in the present section. We
expect that the weights of K and K' in peaks 1 and 2 differ due to $\delta
\neq0$,\ but this should not change the way in which $h_{SDIPS}^{c}$\ shifts
the conductance peaks from $P$\ to $AP$.

In future experiments, it would be interesting to obtain continuous data on a
larger $V_{g}$-range, in order to check that the shape of the $MR(V_{g})$
pattern depends on the occupation of the dot.\ This would also allow to study
the gate voltage dependence of $h_{SDIPS}^{c}$. It would also be interesting
to engineer contacts with ferromagnetic insulators or highly polarized
ferromagnets in order to observe the SDIPS-induced giant $MR$ effect. Note
that although we have considered here the limit $k_{B}T\lesssim\Gamma
_{d\sigma}^{L}+\Gamma_{d\sigma}^{R}$, a strong enough SDIPS should also affect
the behavior of the quantum dot in the sequential tunneling limit $k_{B}%
T\gg\Gamma_{d\sigma}^{L}+\Gamma_{d\sigma}^{R}$, through an analogous mechanism.

\section{Conclusion}

\label{sec:conclusion}

Using an Anderson model, we have studied the behavior of a quantum dot
connected to ferromagnetic leads through spin-active interfaces. The spin
activity of the interfaces makes easier negative magnetoresistance
($\mathrm{MR}$) effects and can even lead to a giant $\mathrm{MR}$ with a sign
oscillating with the gate voltage of the dot. Due to Coulomb blockade, the
$\mathrm{MR}$ versus gate voltage pattern cannot be identical for all
conductance peaks. It is nevertheless possible to account for the
$\mathrm{MR}$ data measured by Ref. \onlinecite{Sahoo} in single-wall carbon
nanotubes by taking into account the $K-K^{\prime}$ orbital degeneracy
commonly observed in those systems.

\section*{Acknowledgments}

A.C. acknowledges discussions with B.~Dou\c{c}ot, T.~Kontos, G.~Montambaux and
I.~Safi. This work was supported by grants from R\'{e}gion Ile-de-France, the
SRC/ERC program (R11-2000-071), the KRF Grant (KRF-2005-070-C00055), and the
SK Fund.


\begin{thebibliography}{99}
\bibitem{Prinz}G. Prinz, Science \textbf{282}, 1660 (1998).

\bibitem{Datta}S. Datta and B. Das, Appl. Phys. Lett. \textbf{56}, 665 (1990).

\bibitem{Schapers}Th. Sch\"{a}pers, J. Nitta, H. B. Heersche, and H.
Takayanagi, Phys. Rev. B \textbf{64}, 125314 (2000); S. Krompiewski, R.
Guti\'{e}rrez, and G. Cuniberti, Phys. Rev. B \textbf{69,} 155423 (2004).

\bibitem{FNF}A. Brataas, Y.V. Nazarov, and G. E. W. Bauer, Phys. Rev. Lett.
\textbf{84}, 2481 (2000); D.H. Hernando, Y.V. Nazarov, A. Brataas, and G.E.W.
Bauer, Phys. Rev. B \textbf{62}, 5700 (2000); A. Brataas, Y.V. Nazarov and
G.E.W. Bauer, Eur. Phys. J. B \textbf{22}, 99 (2001).

\bibitem{Luttinger}L. Balents and R. Egger, Phys. Rev. Lett. \textbf{85}, 3464
(2000); Phys. Rev. B \textbf{64}, 035310 (2001).

\bibitem{Wetzels}W. Wetzels, G. E. W. Bauer, and M. Grifoni, Phys. Rev. B
\textbf{72}, 020407(R) (2005).

\bibitem{Ciuti}C. Ciuti, J.P. McGuire and L.J. Sham, Phys. Rev. Lett. 89,
156601 (2002), J.P. McGuire, C. Ciuti and L.J. Sham, cond-mat/0302088.

\bibitem{ReviewBraatas}A. Brataas, G. E.W. Bauer, and P. J. Kelly, Phys. Rep.
\textbf{427, }157 (2006).

\bibitem{SF}A. Millis, D. Rainer, and J. A. Sauls, Phys. Rev. B \textbf{38},
4504 (1988); M. Fogelstr\"{o}m, \textit{ibid.} \textbf{62}, 11812 (2000); J.C.
Cuevas and M. Fogelstr\"{o}m, \textit{ibid.} \textbf{64}, 104502 (2001); N.M.
Chtchelkatchev, W. Belzig, Y.V. Nazarov, and C. Bruder, JETP Lett.
\textbf{74}, 323 (2001); D. Huertas-Hernando,Y.V. Nazarov, and W. Belzig,
Phys. Rev. Lett. \textbf{88}, 047003 (2002); J. Kopu, M. Eschrig, J. C.
Cuevas, and M. Fogelstr\"{o}m, Phys. Rev. B \textbf{69}, 094501 (2004); E.
Zhao, T. L\"{o}fwander, and J. A. Sauls, \textit{ibid.} \textbf{70}, 134510 (2004).

\bibitem{Tokuyasu}T. Tokuyasu, J. A. Sauls and D. Rainer, Phys. Rev. B
\textbf{38}, 8823 (1988).

\bibitem{SF2005}A. Cottet and W. Belzig, Phys. Rev. B \textbf{72}, 180503(R) (2005).

\bibitem{Tedrow}P. M. Tedrow, J. E. Tkaczyk and A. Kumar, Phys. Rev. Lett.
\textbf{56}, 1746 (1986).

\bibitem{Takis}T. Kontos, M. Aprili, J. Lesueur, and X. Grison, Phys. Rev.
Lett. \textbf{86}, 304 (2001); T. Kontos, M. Aprili, J. Lesueur, F. Gen\^{e}t,
B. Stephanidis, and R. Boursier, \textit{ibid.} \textbf{89}, 137007 (2002).

\bibitem{wire2005}A. Cottet, T. Kontos, W. Belzig, C. Sch\"{o}nenberger and C.
Bruder, Europhys. Lett. \textbf{74}, 320 (2006).

\bibitem{comment}Very recently, Man \textit{et al}.~\cite{Morpurgo} have
reported a spin-dependent transport experiment through a SWNT connected to
collinearly-polarized ferromagnetic leads in the resonant tunneling regime.
The data of this Ref. are interpreted following the lines in Ref.~\onlinecite
{wire2005}. In this particular experiment, the SDIPS has been found to be
vanishing. This is probably due to the fact that, considering the strong
values of $T_{n}$ in this experiment, the effects of SDIPS on the
$\mathrm{MR}(V_{g})$ curves are too weak to be resolved in the actual experiment.

\bibitem{Morpurgo}H.T. Man, I.J.W. Wever and A.F. Morpurgo, cond-mat/0512505.

\bibitem{Sahoo}S. Sahoo, T. Kontos, J. Furer, C. Hoffmann, M. Graber, A.
Cottet and C. Sch\"{o}nenberger, Nature Phys. \textbf{1}, 99 (2005).

\bibitem{question}In Ref. \onlinecite{Sahoo}, the experimental data are
compared with a non-interacting theory which corresponds to the low
transmission limit of Ref. \onlinecite{wire2005}.

\bibitem{expCB}S. J. Tans, M. H. Devoret, J. A. Groeneveld and C. Dekker,
Nature \textbf{394}, 761 (1998).

\bibitem{Sapmaz}S. Sapmaz, P. Jarillo-Herrero, J. Kong, C. Dekker, L. P.
Kouwenhoven, and H. S. J. van der Zant, Phys. Rev. B\textbf{\ 71}, 153402 (2005).

\bibitem{CB}J. Barnas and A. Fert, Phys. Rev. Lett. \textbf{80}, 1058 (1998);
A. Braatas, Yu. V. Nazarov, J. Inoue and G. E. W. Bauer, Phys. Rev. B
\textbf{59}, 93 (1999); H. Imamura, S. Takahashi and S. Maekawa, Phys. Rev. B
\textbf{59}, 6017 (1999); B. R. Bulka, Phys. Rev. B \textbf{62}, 1186 (2000);
A. Cottet, W. Belzig and C. Bruder, Phys. Rev. Lett. \textbf{92}, 206801
(2004), Phys. Rev. B \textbf{70}, 115315 (2004); H.-F. Mu, G. Su, and Q.-R.
Zheng, Phys. Rev. B \textbf{73}, 054414 (2006).

\bibitem{Braun}M. Braun, J. Konig and J. Martinek, Phys. Rev. B \textbf{70},
195345 (2004).

\bibitem{Weymann}I. Weymann, J. K\"{o}nig, J. Martinek, J. Barnas and G.
Schon, Phys. Rev. B \textbf{72} 115334 (2005).

\bibitem{Kondo}N. Sergueev, Q.-F. Sun, H. Guo, B. G. Wang, and J. Wang, Phys.
Rev. B \textbf{65}, 165303 (2002); J. Martinek, M. Sindel, L. Borda, J.
Barnas, J. K\"{o}nig, G. Sch\"{o}n, and J. von Delft, Phys. Rev. Lett.
\textbf{91}, 247202 (2003); M.-S. Choi , D. Sanchez and R. Lopez, Phys. Rev.
Lett. \textbf{92}, 056601 (2004); Y. Utsumi, J. Martinek, G. Sch\"{o}n, H.
Imamura and S. Maekawa, Phys. Rev. B \textbf{71}, 245116 (2005); J. Martinek,
M. Sindel, L. Borda, J. Barnas, R. Bulla, J. K\"{o}nig, G. Sch\"{o}n, S.
Maekawa and J. von Delft, Phys. Rev. B \textbf{72}, 121302(R) (2005); R.
Swirkowicz , M. Wilczynski, M. Wawrzyniak and J. Barnas, Phys. Rev. B
\textbf{73}, 193312 (2006).

\bibitem{Martinek}J. Martinek, Y. Utsumi, H. Imamura, J. Barnas, S. Maekawa,
J. Konig and G. Schon, Phys. Rev. Lett. \textbf{91}, 127203 (2003).

\bibitem{Luttinger2}C. S. Pe\c{c}a, L. Balents and K. J. Wiese, Phys. Rev. B
\textbf{68}, 205423 (2003).

\bibitem{MFL}H.-F. Mu, G. Su, Q.-R. Zheng, and B. Jin, Phys. Rev. B
\textbf{71}, 064412 (2005).

\bibitem{Liang}W. Liang, M. Bockrath, and H. Park, Phys. Rev. Lett.
\textbf{88}, 126801 (2002);

\bibitem{Jarillo}P. Jarillo-Herrero, J. Kong, H. S. J. van der Zant, C.
Dekker, L. P. Kouwenhoven, and S. De Franceschi, Phys. Rev. Lett. \textbf{94},
156802 (2005).

\bibitem{Moriyama}S. Moriyama, T. Fuse, M. Suzuki, Y. Aoyagi, and K.
Ishibashi, Phys. Rev. Lett.\textbf{\ 94}, 186806 (2005);

\bibitem{Babic}B. Babic and C. Sch\"{o}nenberger, Phys. Rev. B \textbf{70},
195408 (2004);

\bibitem{Babic2}B. Babic, T. Kontos, and C. Sch\"{o}nenberger, Phys. Rev. B
\textbf{70}, 235419 (2004)

\bibitem{Blanter}Ya. M. Blanter and M. B\"{u}ttiker, Phys. Rep. \textbf{336},
1 (2000).

\bibitem{granular}L. F. Schelp, A. Fert, F. Fettar, P. Holody, S. F. Lee, J.
L. Maurice, F. Petroff, and A. Vaures, Phys. Rev. B \textbf{56}, R5747 (1997);
K. Yakushiji, S. Mitani, K.Takanashi, S. Takahashi, S. Maekawa, H. Imamura,
and H. Fujimori Appl. Phys. Lett.\textbf{78}, 515 (2001); L. Zhang, C. Wang,
Y. Wei, X. Liu, and D. Davidovic, Phys. Rev.B \textbf{72}, 155445 (2005).

\bibitem{nanoparticles}M.M. Deshmukh and D. C. Ralph, Phys. Rev. Lett.
\textbf{89}, 266803 (2002).

\bibitem{tubes}K. Tsukagoshi, B. W. Alphenaar, and H. Ago, Nature
\textbf{401}, 572574 (1999); B. Zhao, I.M\"{o}nch, H. Vinzelberg, T. M\"{u}hl
and C. M. Schneider, Appl. Phys. Lett. \textbf{80}, 31443146 (2002).

\bibitem{molecules}A. Pasupathy et al., Science \textbf{306}, 86 (2004).

\bibitem{explanationWetzels}Note that Ref. \onlinecite{Wetzels} has also
introduced an effective exchange field to take into account the SDIPS in an
interacting single electron transistor connected to ferromagnetic leads. These
authors have studied precession effects produced by the SDIPS\ in
non-collinear configurations. They did not consider spin-dependent resonance
effects because this is not relevant in the diffusive limit considered in this Ref.

\bibitem{Meir2}Y. Meir and N.S. Wingreen, Phys. Rev. Lett. \textbf{68}, 2512 (1992).

\bibitem{Meir}Y. Meir, N.S. Wingreen and P.A. Lee, Phys. Rev. Lett.,
\textbf{66,} 3048 (1991).

\bibitem{note SA}Note that in the limit considered in this article, the spin
accumulation concept of Ref. \onlinecite{FNF} simply corresponds to having a
finite average spin on the quantum dot. This feature is intrinsically taken
into account in our treatment.

\bibitem{RelModel}For $U=0$, the conductance given by Eqs. (\ref{cond}) and
(\ref{Gtot}) can be perfectly mapped onto the interactionless conductance
found in Ref.\onlinecite{wire2005} for $T_{l}^{c,\sigma}\ll1$ and a level
$E_{d\sigma}^{c}$ close to resonance, using $E_{d\sigma}^{c}=\xi_{d\sigma}$
and $T_{l}^{c,\sigma}=\pi N_{F}^{M}\Gamma_{d\sigma}^{l}=2\pi N_{F}%
^{M}\left\vert \Sigma_{d\sigma}^{0}\right\vert $.

\bibitem{Soulen}R. J. Soulen Jr., J. M. Byers, M. S. Osofsky, B. Nadgorny, T.
Ambrose, S. F. Cheng, P. R. Broussard, C. T. Tanaka, J. Nowak, J. S. Moodera,
A. Barry, and J. M. D. Coey, Science \textbf{282}, 85, (1998).

\bibitem{TKferro}From Ref.\cite{Martinek}, for $\left\vert P_{L(R)}\right\vert
=0.2$, the Kondo temperature $T_{K}$ of the circuit is very close to the
$T_{K}$ corresponding to $\left\vert P_{L(R)}\right\vert =0 $ in both the $P$
and $AP$ configurations.

\bibitem{Aleiner}I.L. Aleiner, P.W. Brouwer and L.I. Glazman, Phys. Rep.
\textbf{358}, 309 (2002).

\bibitem{PC}T. Kontos has communicated us two extra points at $V_{g}\sim
4.417$~\textrm{V}.

\bibitem{pointout}We point out that this splitting effect is unrelated to
$h_{SDIPS}^{c}$. This is particularly clear in the limit of no tunneling
$\left|  t_{d\sigma}^{k}\right|  \rightarrow0$, in which the renormalization
of the levels by the $\Sigma_{d\sigma,d^{\prime}\sigma^{\prime}}^{i,n}$ terms
vanishes whereas $h_{SDIPS}^{c}$ persists.

\bibitem{CorrExch}The function $G_{d\sigma}(\omega)$ is resonant at
$\hbar\omega=\xi_{d\sigma}^{S}=\xi_{d\sigma}+\operatorname{Re}[\Sigma
_{\overline{\sigma}}^{S}]$ and $\hbar\omega=\xi_{d\sigma}^{D}=\xi_{d\sigma
}+\operatorname{Re}[\Sigma_{\overline{\sigma}}^{D}]$. Interestingly, for
$U\gg\Gamma_{\sigma}$, $\xi_{d\uparrow}=\xi_{d\downarrow}=\xi$, $\Gamma
_{L}=\Gamma_{R}=\Gamma$ and $P_{L}=P_{R}=P$, a simplified expression of the
resonance spin-splitting caused by the $\Sigma_{d\sigma,d^{\prime}%
\sigma^{\prime}}^{i,n}$ terms can be obtained by evaluating $\operatorname{Im}%
[\Sigma_{\overline{\sigma}}^{S}]$ at $\hbar\omega=\xi_{d\sigma}$ and
$\operatorname{Im}[\Sigma_{\overline{\sigma}}^{D}]$ at $\hbar\omega
=\xi_{d\sigma}+U$. \ This leads to $\xi_{d\uparrow}^{S}-\xi_{d\downarrow}%
^{S}=\xi_{d\uparrow}^{D}-\xi_{d\downarrow}^{D}=g\mu_{B}h_{U}^{c}=(P\Gamma
/\pi)\int^{^{\prime}}d\omega\lbrack\{f(\omega)/(\omega-\xi-U)\}+\{(1-f(\omega
))/(\omega-\xi)\}]$ where the prime represents Cauchy's principal value.
Interestingly, this expression perfectly matches with Eq. (3.9) of Ref.
\onlinecite{Braun}. This Ref. studies the precession effect produced by
$h_{U}^{c}$ in a quantum dot with ferromagnetic leads polarized in
non-collinear directions, in the limit\ $\Gamma_{\sigma}\gg k_{B}T$. The
system is described with Eqs. which neglect $h_{U}^{c}$ in the collinear case.

\bibitem{SU4exp}P. Jarillo-Herrero, J. Kong, H. S. J. van der Zant, C. Dekker,
L. P. Kouwenhoven and S. De Franceschi , Nature \textbf{434}, 484 (2005).

\bibitem{SU4th}M.-S. Choi, R. L\'{o}pez and R. Aguado, PRL \textbf{95}, 67204 (2005).

\bibitem{IfNot}We expect that the effect of the SDIPS is only quantitatively
modified when these processes occur.

\bibitem{multi}P. Pals and A. MacKinnon, J. Phys. Condens. Matter \textbf{8},
5401 (1996), J.J. Palacios, L. Liu and D. Yoshioka, Phys. Rev. B \textbf{55},
15735 (1997).

\bibitem{MoreSym}Note that it is possible to decrease the up spins
contribution in peak 3 and to have a $MR(V_{g})$ pattern similar for peaks 3
and 4 by using values of $P_{L(R)}$ much smaller than in Fig.\ref{Figure3}
(e.g. $P_{L(R)}=0.05$).

\bibitem{stray}B. W. Alphenaar, K. Tsukagoshi and M. Wagner, J. Appl. Phys.
\textbf{89}, 6863 (2001).

\bibitem{bockrath}W. Liang, M. Bockrath, D. Bozovic, J.H. Hafner, M. Tinkham
and H. Park, Nature, \textbf{411}, 665 (2001).

\bibitem{note}See Fig. 1 of Ref. \onlinecite{wire2005}

\bibitem{Oreg}Y. Oreg, K. Byczuk and B.I. Halperin, Phys. Rev. Lett.
\textbf{85}, 365 (2000).
\end{thebibliography}
\end{document}